\documentclass[12pt,preprint]{emulateapj}

\newcommand{\jcap}{Journal of Cosmology and Astroparticle Physics}

\usepackage{graphicx}
\usepackage{amsmath}
\usepackage{hyperref}
\usepackage{color}

\hypersetup{
    colorlinks=true,
    linkcolor=red,
    citecolor=blue,
    urlcolor=magenta,
}

\begin{document}

\title{A model-independent determination of the Hubble constant from lensed quasars and supernovae using Gaussian process regression}
\author{Kai Liao$^{1}$}
\author{Arman Shafieloo$^{2,3}$}\email{shafieloo@kasi.re.kr}
\author{Ryan E. Keeley$^{2}$}
\author{Eric V. Linder$^{2,4,5}$}
\affil{
$^1$ {School of Science, Wuhan University of Technology, Wuhan 430070, China}\\
$^2$ {Korea Astronomy and Space Science Institute, Daejeon 34055, Korea}\\
$^3$ {University of Science and Technology, Daejeon 34113, Korea}\\
$^4$ {Berkeley Center for Cosmological Physics and Berkeley Lab, University of California, Berkeley, CA 94720, USA}\\
$^5$ {Energetic Cosmos Laboratory Nazarbayev University, Nur-Sultan, Kazakhstan 010000}
}

\begin{abstract}
Strongly lensed quasar systems with time delay measurements provide ``time
delay distances", which are a combination of three angular diameter distances and serve as powerful tools to determine the Hubble constant $H_0$. However, current results often rely on the assumption of the $\Lambda$CDM model.
Here we use a model-independent method based on Gaussian process to directly constrain the value of $H_0$.
By using Gaussian process regression, we can generate posterior samples of unanchored supernova distances independent of any cosmological model and anchor them
with strong lens systems. The combination of a supernova sample with large
statistics but no sensitivity to $H_0$ with a strong lens
sample with small statistics but $H_0$ sensitivity gives
a precise $H_0$ measurement
without the assumption of any cosmological model. We use four well-analyzed lensing systems from the state-of-art lensing program H0LiCOW and the Pantheon supernova compilation in our analysis.
Assuming the Universe is flat, we derive the constraint  $H_0=72.2 \pm 2.1\,$km/s/Mpc, a precision of $2.9\%$. Allowing for cosmic curvature with a prior of $\Omega_{k}=[-0.2,0.2]$, the constraint becomes $H_0=73.0_{-3.0}^{+2.8}\,$km/s/Mpc.
\end{abstract}

\section{Introduction}

The standard cosmological model of $\Lambda$CDM has achieved great successes in explaining a wide array of cosmological observations, including the expansion history from distances of Type Ia supernovae (SNe) and baryon acoustic oscillations
(BAO), the cosmic growth history from galaxy surveys, and the
cosmic microwave background (CMB). However, there are some
discrepancies, the clearest of which is
the value of the Hubble constant $H_0$ as measured locally,
from SNe Ia calibrated by Cepheid variable stars \citep{2019ApJ...876...85R} or the
tip of the red giant branch of stars \citep{Freedman1907.05922}, and as derived cosmologically,
from CMB \citep{Planck18} or BAO with or without CMB or indeed
any early universe information \citep{1707.06547,1811.02376,1906.11628}.

Note that the cosmological estimation of $H_0$ is not a
direct measurement but a derivation as one of a number of
cosmological model parameters. Usually the model assumed is
the $\Lambda$CDM when inferring $H_0$. Therefore, the discrepancy reveals either new physics beyond the standard model or unknown systematic errors in the observations. To better understand this tension problem, one may use new cosmological approaches to determine $H_0$ model-dependently or even model-independently. For example, gravitational wave events from inspiraling compact objects plus their  electromagnetic counterparts are very promising for an
alternate local direct measurement~\citep{Abbott2017}.

Strong gravitational lensing systems with time delays offer
a method for cosmological determination of $H_0$ that is
partly direct (time delays are proportional to $1/H_0$)
and partly derived from a cosmological parameter fit, but
independent of both the local and the early universe.
A typical system consists of a lensed quasar at cosmological distance, lensed by a foreground elliptical galaxy, forming multiple images of the active galactic nucleus (AGN) and the arcs of its host galaxy. With years of monitoring on the AGN light curves, one can measure the time delay between any two images corresponding to different paths, due to the geometric and Shapiro effects following the Fermat principle. The time delay thus depends on both the geometry of the Universe and the gravity field of the lens galaxy (lensing potential).

With ancillary data to measure the lensing potential, for example  high-resolution imaging, stellar dynamic measurements, and line-of-sight
environment measurements, we can measure the geometry of the Universe in terms of the ``time delay distance" $D_{\Delta t}$. This is a ratio of three angular diameter distances, and depends on $H_0$ and also other cosmological parameters. In this way strong lensing has been used to determine $H_0$. The lensing program H0LiCOW \citep{Suyu2017} with the Hubble Space Telescope can measure $D_{\Delta t}$ with several percent precision for each system, from percent level measurements of both time delays and lens modeling.

With only four well-measured systems, they constrained $H_0$ with $3\%$ precision in a flat $\Lambda$CDM model~\citep{Birrer2019}, and looser precision in other cosmological models where the dark energy equation of state is allowed to differ from $-1$. A more ambitious aim~\citep{Suyu2017} is to measure $H_0$ within $1\%$ uncertainty such that the current tension may be investigated with high statistical significance. Though the measured $H_0$ would be very important for the community, it is worth noting that these measurements rely on assuming certain cosmological models. In \cite{Taubenberger2019},
they found the results could be stable with respect to different models when combining strong lensing with SNe Ia as another
cosmological probe\footnote{One might wonder whether some of this could possibly be due to using JLA SNe data in some tension with lensing data, such that their combination
pulled in on $\Lambda$CDM where all the models considered were similar.
\citet{Wong2019} did find that using Pantheon data increased the spread in the mean
$H_0$ value from flat $\Lambda$CDM to flat $w_0w_a$CDM from 0.6 for JLA to 1.4 for
Pantheon, which is still not significant compared to the uncertainties.
}.

The Inverse distance ladder method~\citep{Aubourg2015,Cuesta2015} provides a more model-independent way to infer $H_0$.
The idea is to anchor the relative distances from SNe Ia with the absolute distance measurements from other cosmological approaches.
In this work, we propose a model-independent approach to determining $H_0$ using a Gaussian process (GP) and by anchoring the SNe Ia with strong lensing. This combines the strengths of each technique: the large sample of SNe constrains the
individual distances, freeing the smaller sample of strong lenses to lock down $H_0$ from the distance ratio. Having $H_0$ determined in this way could be more direct and informative for understanding the Hubble tension problem since the results are not biased by any model or parametric assumption -- and are independent of both
the local universe and the very high redshift, early
universe. This gives a new angle on the problem.

We should note that a recent work by \cite{Collett2019} 
used SNe Ia and strong lensing to determine $H_0$ by implementing fourth order polynomial fitting to the supernovae data. Comparison of our results will be interesting,
especially because using parametric approaches such as polynomial fitting to different cosmology data can be prone to issues such as instabilities, e.g.\ see
\citep{2004JCAP...09..007J,2006MNRAS.366.1081S,2007MNRAS.380.1573S,Holsclaw,Holsclaw1,Holsclaw2,Shaf2012,ShafKimLind}.
It is therefore useful to crosscheck, especially as using model-independent or
non-parametric reconstruction approaches, as we do in this work, reduces potential
bias from the forms of the parametric or model assumptions.

This Letter is organized as follows: In Section 2 we introduce the time delay cosmology and the latest lensing data.
Then we combine lensing and SNe Ia to give a model-independent constraint on $H_0$ in Section 3. We summarize and make
discussions in Section 4.
Throughout this Letter, we use the natural units of c=G=1 in all equations. $H_0$ in units of km/s/Mpc will be recovered in the results.

\section{Time delay distances from lensing}

According to the theory of strong lensing, the time delay between two images of the AGN is determined by
both the geometry of the Universe and the gravity field of the lens galaxy through:
\begin{equation}
\Delta t=D_{\Delta t}\Delta \phi(\boldsymbol{\xi}_{lens}),\label{Dt}
\end{equation}
where
$\Delta\phi=[(\boldsymbol{\theta}_A-\boldsymbol{\beta})^2/2-\psi(\boldsymbol{\theta}_A)-(\boldsymbol{\theta}_B-\boldsymbol{\beta})^2/2+\psi(\boldsymbol{\theta}_B)]$
is the difference of Fermat potentials at two images,
$\boldsymbol{\theta}_A$ and $\boldsymbol{\theta}_B$ denote the angular positions of the images while
$\boldsymbol{\beta}$ denotes the position of the source (supposing it is unlensed).
$\psi$ is the two-dimensional lensing potential via the Poisson equation $\nabla^2\psi=2\kappa$,
where $\kappa$ is the surface mass density of the lens
in units of critical density $\Sigma_{\mathrm{crit}}=D_s/(4\pi D_lD_{ls})$.
Assuming a lens model, $\Delta \phi$ is determined by the parameters $\boldsymbol{\xi}_{lens}$ therein.
$D_{\Delta t}$ is the ``time delay distance" formed from three angular diameter distances:
\begin{equation}
D_{\Delta t}=(1+z_l)\frac{D_lD_s}{D_{ls}},\label{defDt}
\end{equation}
where $l,s$ stands for lens and source. Note that $D_{\Delta t}$ is primarily sensitive to $H_0$, providing
a powerful and independent way to determine it.

The $\Delta t$ can be measured by comparing the light curve shift of the two AGN images. With current techniques and the
quality of the light curves, the precision of $\Delta t$ can be up to percent levels
(see, e.g., \citet{Tewes2013} among many others). The upcoming LSST will discover
thousands of lensed quasars, some of which will have well-measured light curves. The Time Delay Challenge~\citep{Liao2015} showed about
400 systems will have robust time delay measurements with average precision $3\%$, making time delay cosmography very promising.
Meanwhile, the Fermat potentials can be measured by high-resolution imaging from space telescopes or
ground-based adaptive optics, together with the stellar
dynamics and the structure along the line-of-sight. The precision of $\Delta \phi$ is comparable with that of $\Delta t$,
resulting in few percent level determination of $D_{\Delta t}$ for each system.

The state-of-art lensing project H0LiCOW aims at measuring $H_0$ with precision $\lesssim1\%$ based on
a small sample of well-observed lenses in the near future. Currently, under a flat $\Lambda$CDM model, they get a result with $3\%$ precision including systematics, from only 4 lenses: RXJ1131-1231, HE 0435-1223, B1608+656 and SDSS 1206+4332. Table~\ref{redshifts} lists the lens and source redshifts for these systems.
The data from another two lenses should also be released soon~\citep{Chen2019,Rusu2019,Wong2019}.
The posteriors of the time delay distances for the four lenses are given in the H0LiCOW papers
and  website\footnote{http://www.h0licow.org}.
For the first three of them, the posterior probability distributions of $D_{\Delta t}$ are described by the following analytic fit:
\begin{equation}
\small
P(D_{\Delta t})=\frac{1}{\sqrt{2\pi}(x-\lambda_D)\sigma_D}\,\exp\left\{-\frac{[\ln(x-\lambda_D)-\mu_D]^2}{2\sigma_D^2}\right\},
\end{equation}
where $x=D_{\Delta t}/(1\,{\rm Mpc})$ and the parameters $(\lambda_D,\sigma_D,\mu_D)$ can be found
in Table~3 of \citet{Bonvin2017}. For the fourth lens SDSS 1206+4332, the posterior is given in the form of
the Markov chain Monte Carlo (MCMC)~\citep{Birrer2019}.

\begin{table}\centering
\begin{tabular}{cllc}
\hline\hline
Order & Name & $z_L$ & $z_S$ \\
\hline
1 & RXJ1131-1231 & 0.295 & 0.654 \\
2 & HE 0435-1223 & 0.4546 & 1.693 \\
3 & B1608+656 & 0.6304 & 1.394  \\
4 & SDSS 1206+4332 & 0.745 & 1.789 \\
\hline\hline
\end{tabular}
\caption{Lens and source redshifts for the four strong lens systems ordered by distance (see Taubenberger et al. 2019 and the references therein).
}\label{redshifts}
\end{table}

However, since the angular distances depend on the cosmological model, the time delay distances also
change in different cosmological models.
The value of $H_0$ can then vary considerably between models.
\cite{Taubenberger2019} claim to achieve stable results for $H_0$, i.e.\ insensitive to cosmological model, by using strong lensing to
anchor Type Ia supernova data. However, that analysis was still within a set of cosmological models.
Furthermore, it used the JLA supernovae data, which mildly prefer a dark energy equation of state $w>-1$ (see Fig. 16 of \cite{Betoule14}),
while strong lenses mildly prefer $w<-1$ (\cite{Wong2019}). The combination therefore may lie close to the cosmological constant
$w=-1$ where the models they consider are equivalent. Thus, it is useful to try a model independent
analysis for $H_0$, with the more recent Pantheon supernovae data set.

\section{Methodology and results}

SNe Ia observations led to the discovery of cosmic acceleration. They are incisive probes of
cosmology through determining the shape of the cosmic distance-redshift relation. That is,
they determine distances in a relative sense, but the absolute distance is convolved with a
combination of the absolute magnitude of SNe Ia and the Hubble constant. The SNe data can be
combined with (``anchored by'') an absolute distance probe (for example Cepheid variable stars,
the tip of the red giant branch stars, or gravitational wave
sirens in the local universe) to form an absolute distance probe.
In addition, absolute distance probes at cosmological distances, e.g.\ strong lens systems, can anchor SN. This greatly benefits the leverage of the absolute
probe in constraining cosmology since SNe tend to be much more numerous as well as very good
probes of dark energy properties, mapping a wide range of cosmic
expansion history. Thus the combination of strong lensing time delays and SNe
can determine both $H_0$ and the cosmic expansion history.

\subsection{Data and Method}
For the SNe data we use the most recent and largest data set, the Pantheon compilation \citep{pantheon}.
To combine the Pantheon SNe and the H0LiCOW strong lenses datasets, we generate a posterior sampling of the $H_0$-independent quantity $H_0 D^L(z)$ from the Pantheon dataset. To do the posterior sampling in a manner independent of a cosmological model, we use GP regression~\citep{Holsclaw,Holsclaw1,Holsclaw2,ShafKimLind,ShafKimLind2}. The GP regression used here is based on the \texttt{GPHist} code \citep{GPHist} first used in \cite{Keeley0}.  GP regression works by generating a family of functions over an infinite dimensional function space as determined by a kernel.
We use
\begin{equation}
    \langle \gamma(z_1)\gamma(z_2) \rangle = \sigma_f^2 \, \exp\{-[s(z_1)-s(z_2)]^2/(2\ell^2)\},
\end{equation}
where $\sigma_f$ and $\ell$ are hyperparameters, respectively characterizing the amplitude of
variations with redshift and their correlation scale. The hyperparameters play important roles for
both physical insight and error control, and must be fit for not fixed. The priors on the GP hyperparameters are scale-invariant i.e. flat in the log of the hyperparameters.  Since the dimmensionality of the hyperparameters is small, we directly integrate over the hyperparameters.

We then use GP on the SNe data to generate expansion histories $H(z)/H_0$ where $\gamma(z) = \ln([H^{\rm fid}(z)/H_0]/[H(z))/H_0])$. Here $H^{\rm fid}(z)/H_0$ is taken to be the best fit $\Lambda$CDM model for the Pantheon data and serves the role of the mean function for GP regression. Such prewhitening is standard practice and extensive tests show the resulting median inference does not depend on the details of the mean function \citep{ShafKimLind,ShafKimLind2,Aghamousa2017}.
With $H(z)/H_0$ in hand, we can calculate the unanchored SNe luminosity distances

\begin{equation}
    H_0 D^L(z) = (1+z) \Omega_k^{-1/2}\sinh\left[\Omega_k^{1/2} \int^z_0 dz'/[H(z')/H_0]\right] \ ,
\end{equation}


(where $\sinh$ is a complete function valid for all signs of $\Omega_k$) and any corresponding angular diameter distances $H_0 D^A=H_0 D^L/(1+z)^2$ we will need later for
the strong lensing systems, where $\Omega_k=1-\Omega_{\rm total}$ is the curvature energy density in units of the critical density.
The likelihood of how well these SNe distances fit the Pantheon data are then used as weights in randomly selecting 1000 samples used for $D^A$ in the strong lens analysis.

\begin{figure}
 \includegraphics[width=\columnwidth,angle=0]{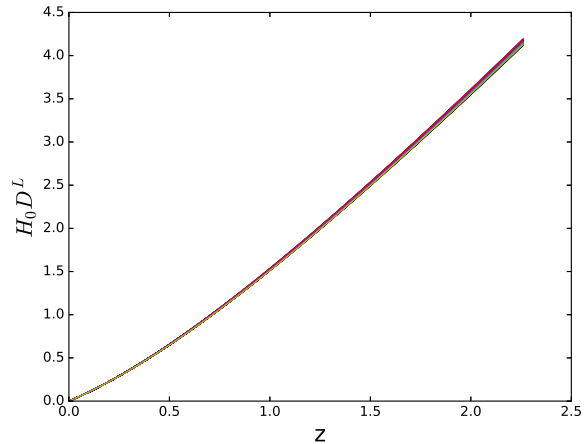}
  \caption{
The unanchored luminosity distance $H_0D^L(z)$ reconstructed
from the SNe data is plotted vs redshift for a representative
sample of the 1000 GP realizations.
  } \label{fig:sn}
\end{figure}

Example GP curves are shown in Fig.~\ref{fig:sn}. The data is dense and precise enough to provide
a well constrained distance-redshift relation.
The spread is about 2\% at $z=1$, and the GP is not constrained
to follow the input mean function -- in fact it deviates from
it by about 1.5\% at $z=1$. Note that the GP covers the full
range of the strong lensing system redshifts so there is no
extrapolation needed.

To summarize the method for constraining $H_0$:
\begin{enumerate}
\item
Draw 1000 unanchored luminosity distance curves $H_0D^L$ from the GP fit to the SNe data, and convert
to unanchored angular diameter distances $H_0D^A$;
\item Evaluate the values of each of the 1000 $H_0D^A$ curves at the lens and source redshifts of the four
strong lens systems to calculate 1000 values of $H_0D_{\Delta t}$ for each system
using $H_0D_{\Delta t}=(1+z_l)(H_0D_l)(H_0D_s)/(H_0D_{ls})$;
\item Compute the likelihood, for each of the 1000 realizations, from the H0LiCOW's $D_{\Delta t}$ data for each lens system for many values of $H_0$;
\item Multiply the four likelihoods to form the full likelihood for each realization, for each value of
$H_0$;
\item Marginalize over the realizations to form the posterior distribution of $H_0$.
\end{enumerate}

Note that to obtain the angular diameter distance between the lens and the source we use the
standard distance relation \citep{Weinberg1972}
\begin{equation}
\begin{split}
D_{ls}&=D_s\sqrt{1+\Omega_{k}(1+z_l)^2(H_0D_l)^2}  \\
&\qquad-\frac{1+z_l}{1+z_s}D_l\sqrt{1+\Omega_{k}(1+z_s)^2(H_0D_s)^2} \ .
\end{split}
\end{equation}
Note that for a spatially flat universe
one simply has
$H_0D^L(z)=(1+z)\int_0^z dz'/[H(z')/H_0]$ and $D_{ls}=D_s-[(1+z_l)/(1+z_s)]\,D_l$.

\subsection{Results}
\begin{figure}
 \includegraphics[width=\columnwidth,angle=0]{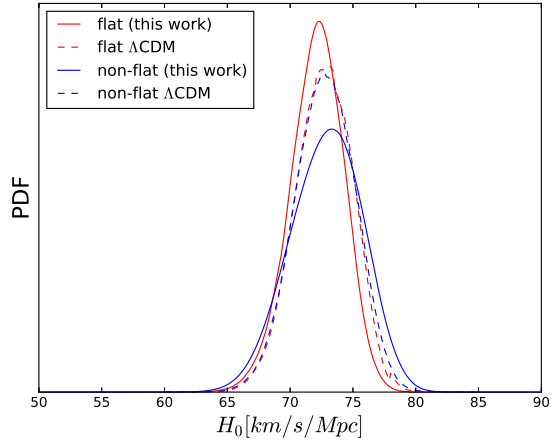}
  \caption{The probability distribution functions (PDFs) of $H_0$ in the cases of flat  and non-flat
  universes. Note that the solid curves labeled ``(this work)'' indicate results from our model independent method while the dashed curves assumed $\Lambda$CDM using strong lensing only~\citep{Taubenberger2019}.
  } \label{H0}
\end{figure}

The final posterior distribution for $H_0$ in the flat universe case is shown in Fig.~\ref{H0}.
Our model-independent constraint is
$H_0=72.2 \pm 2.1\,$ km/s/Mpc (median value plus the $16^{th}$ and $84^{th}$ percentiles around this).
This can be compared to the time delay plus SNe result of
\citet{Taubenberger2019} of $73.1^{+2.1}_{-2.2}$ km/s/Mpc within $\Lambda$CDM or $73.1\pm3.0$ km/s/Mpc within $w_0w_a$CDM, or of
\citet{Wong2019} of $73.6^{+1.6}_{-1.8}$ km/s/Mpc and $75.0^{+2.2}_{-2.3}$ km/s/Mpc respectively. Note that the first set uses JLA SNe rather than
Pantheon and the second set uses six lens systems rather than
four. In addition, the first set uses $D_{\Delta t}$ only while the second uses the combination of $D_{\Delta t}$
and $D_l$.
The main point, however, is that the uncertainties for a
model independent analysis combining strong lensing time delays and a wide ranging
distance probe such as supernovae can be comparable to those assuming
a specific model, while reducing possible bias.

For the case where spatial curvature is included, we
set the flat prior on $\Omega_k=[-0.2,0.2]$  as in \cite{Taubenberger2019}.
The posterior distribution is shown in Fig.~\ref{H0} also, and
the marginalized distributions give $H_0=73.0_{-3.0}^{+2.8}\,$ km/s/Mpc and $\Omega_{k}=0.07_{-0.14}^{+0.09}$.
The weaker constraint is due to the covariance between $H_0$ and $\Omega_{k}$.
While consistent with the flat universe results, note the posterior distribution is more nonGaussian.

\begin{figure*}
    \centering
  \includegraphics[width=0.8\textwidth,angle=0]{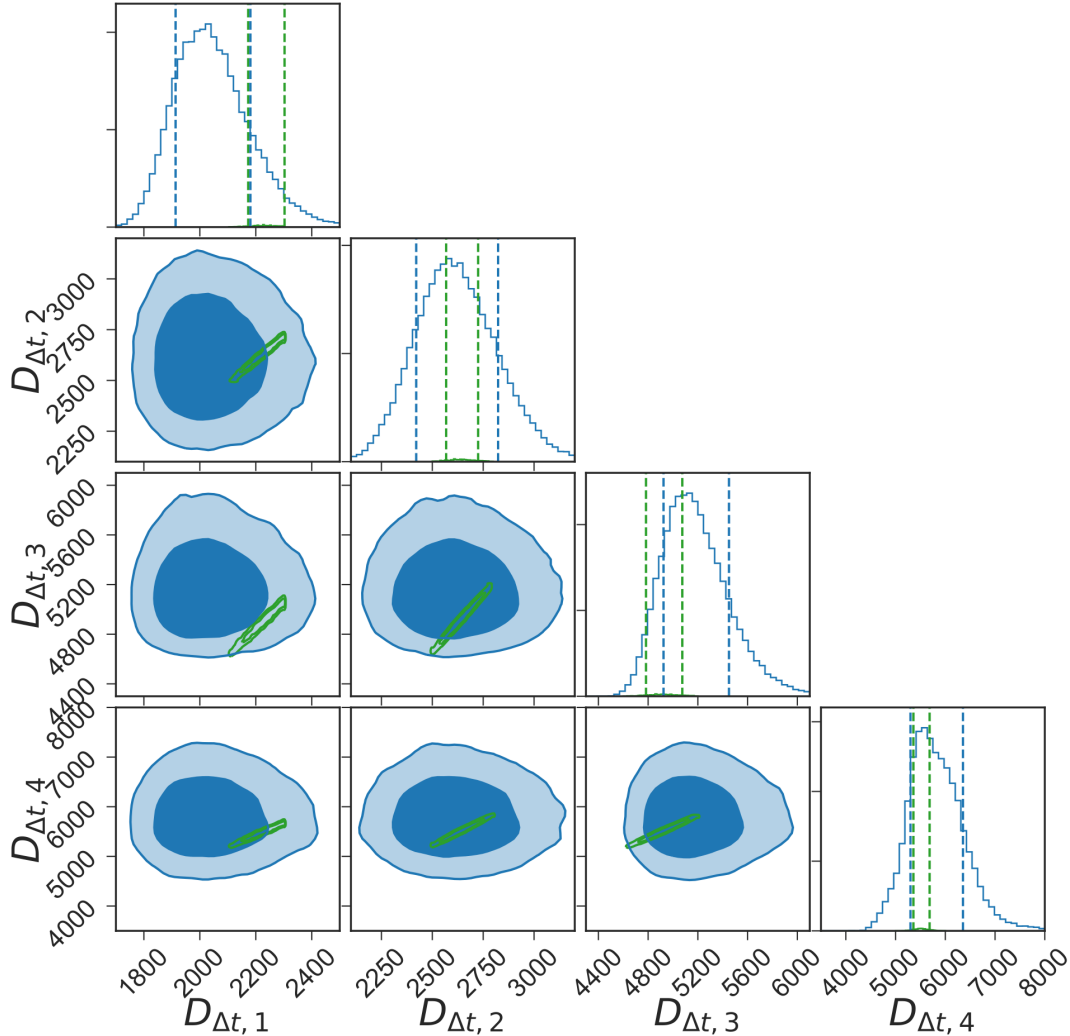}
    \caption{
    2D contours of the likelihoods (68.3\% confidence level (CL) inner, 95.4\% CL outer), and 1D marginalized probability
    distribution functions (68.3\% CL), of the strong lensing time delay distances (blue) and the posterior sampled distances calculated from the GP reconstruction from SNe (green). The units are Mpc. The major axis of the green SNe contours
    corresponds to variation in the value of $H_0$.
    Systems are ordered by time delay distance from lowest to
    highest, with $D_{\Delta t,1}$: RXJ1131-1231, $D_{\Delta t,2}$: HE0435-1223, $D_{\Delta t,3}$: B1608+656, $D_{\Delta t,4}$: SDSS1206+4332.
    }
    \label{fig:consistency2D}
\end{figure*}

When trying to resolve tensions, it is worthwhile checking for the internal
consistency of data combinations used. We explore first consistency between the
time delay distances and SNe distances, and then consistency within
the set of time delay systems, assuming a flat universe.
To check the consistency between the distances from the SNe reconstructions and the time delay distances from the strong lenses, we use the GP fit to the
expansion history derived from the SNe luminosity distances to compute predicted
time delay distances with appropriate $z_l$ and $z_s$. As the SNe luminosity distances are unanchored, the conversion from
$H_0D_{\Delta t}$ to $D_{\Delta t}$ stretches out the joint contours along this degeneracy.
Figure~\ref{fig:consistency2D} shows the results of this consistency check. The distances are indeed consistent, and
hence the combination of the two probes is justified.

\subsection{Consistency}
One could also evaluate the consistency between the best fit
time delay distances from the lensing data and the SNe reconstruction
using the best fit $H_0$ from the combination.
For only four points, a $\chi^2$ has limited significance but we can
mention that $\chi^2=2.28, 0.15, 0.80, 0.24$ for RXJ1131-1231, HE0435-1223, B1608+656, SDSS1206+4332, respectively,
for the data plotted with respect to the GP
distance relation.
We note
that with the systems ordered by increasing time delay distance, one does see
a trend where the lensing data monotonically climbs higher above the relation
predicted by the SNe sample with distance. For such a small lensing data
sample it is difficult to tell whether this reflects a real systematic.
(Note that \citet{Wong2019} shows another monotonic trend, in the derived
$H_0$, with $D_{\Delta t}$, using six strong lens systems). It would be
interesting as data sets get larger to study not only the mean $H_0$ derived,
but whether any trends exist, which could potentially point to systematics
with distance (or other physical characteristics) such as line of sight mass
corrections or stellar dynamics scale effects.

\begin{figure*}
    \centering
    \includegraphics[width=0.4\textwidth,angle=0]{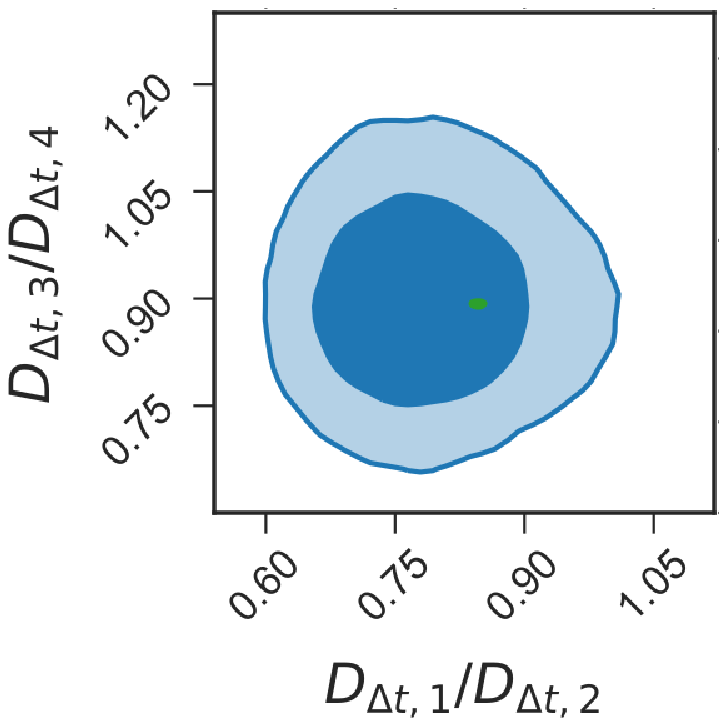}\qquad
  \includegraphics[width=0.4\textwidth,angle=0]{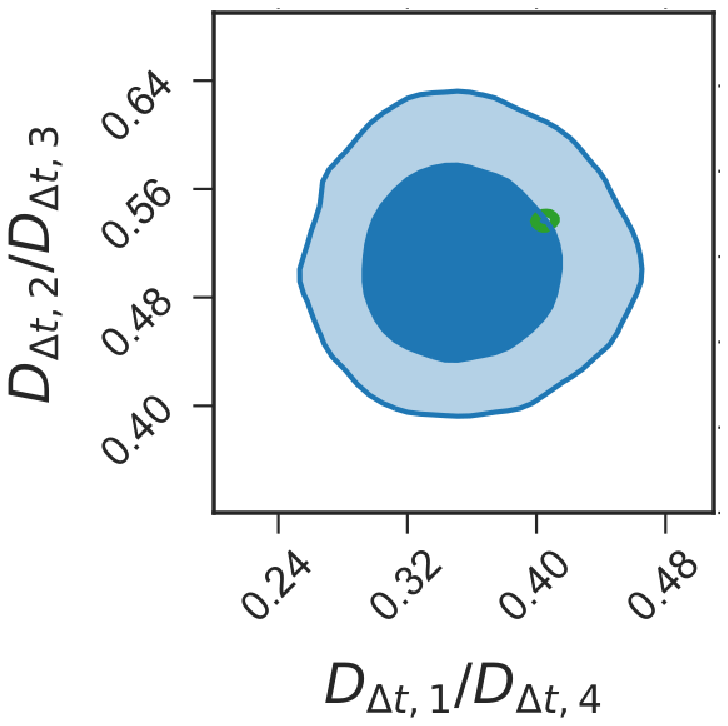}
    \caption{
    The likelihood contours (68.3\%CL inner, 95.4\% CL outer) of the ratios of
    pairs of time delay distances for both the strong lensing data (blue) and the SNe reconstructions (green).
    The left panel shows neighbors: the ratio of the smallest two vs largest two distances; the right panel shows extremes: the ratio of the smallest to largest vs middle two distances.
    }
    \label{fig:ratio}
\end{figure*}

To avoid assuming any value for $H_0$ to employ the SNe data, we can also
consider ratios of time delay distances, which are independent of $H_0$.
Furthermore, to explore the possibility of trends we plot these against each other. Rather than show all 15 plots of pair combinations of the 6
ratios, we show only 2 in Fig.~\ref{fig:ratio}: neighbors -- $D_{\Delta t,1}/D_{\Delta t,2}$ vs
$D_{\Delta t,3}/D_{\Delta t,4}$, and extremes -- $D_{\Delta t,1}/D_{\Delta t,4}$ vs $D_{\Delta t,2}/D_{\Delta t,3}$.

We see that the data and reconstructed distances are consistent
at the 68.3\% confidence level. Most of the apparent trend with
distance seems to be due to the nearest lens, RXJ1131-1231. This
could be simply a statistical fluctuation but note that \cite{1511.03662}
found that the use of a different lens galaxy kinematic prior for this
system could shift the value of $H_0$ from it down by 14\%. In addition,
\citet{Chen2019}, using adaptive optics imaging only, reduced its value of
$H_0$ from $78.2\pm3.4$ km/s/Mpc to $77.0^{+4.0}_{-4.6}$ km/s/Mpc (though not a
significant shift).

\begin{figure*}
    \centering
    \includegraphics[width=0.4\textwidth,angle=0]{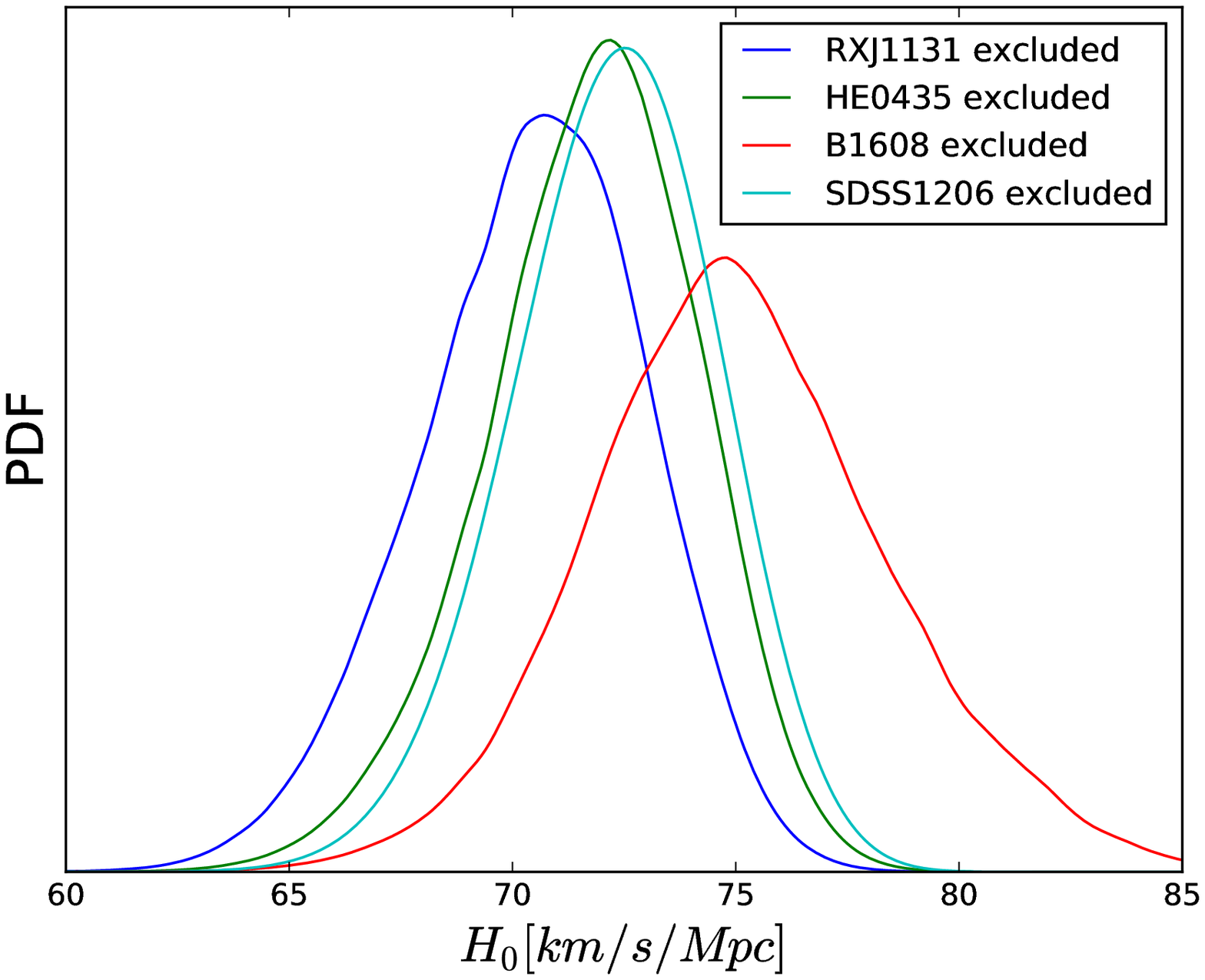}\qquad
  \includegraphics[width=0.4\textwidth,angle=0]{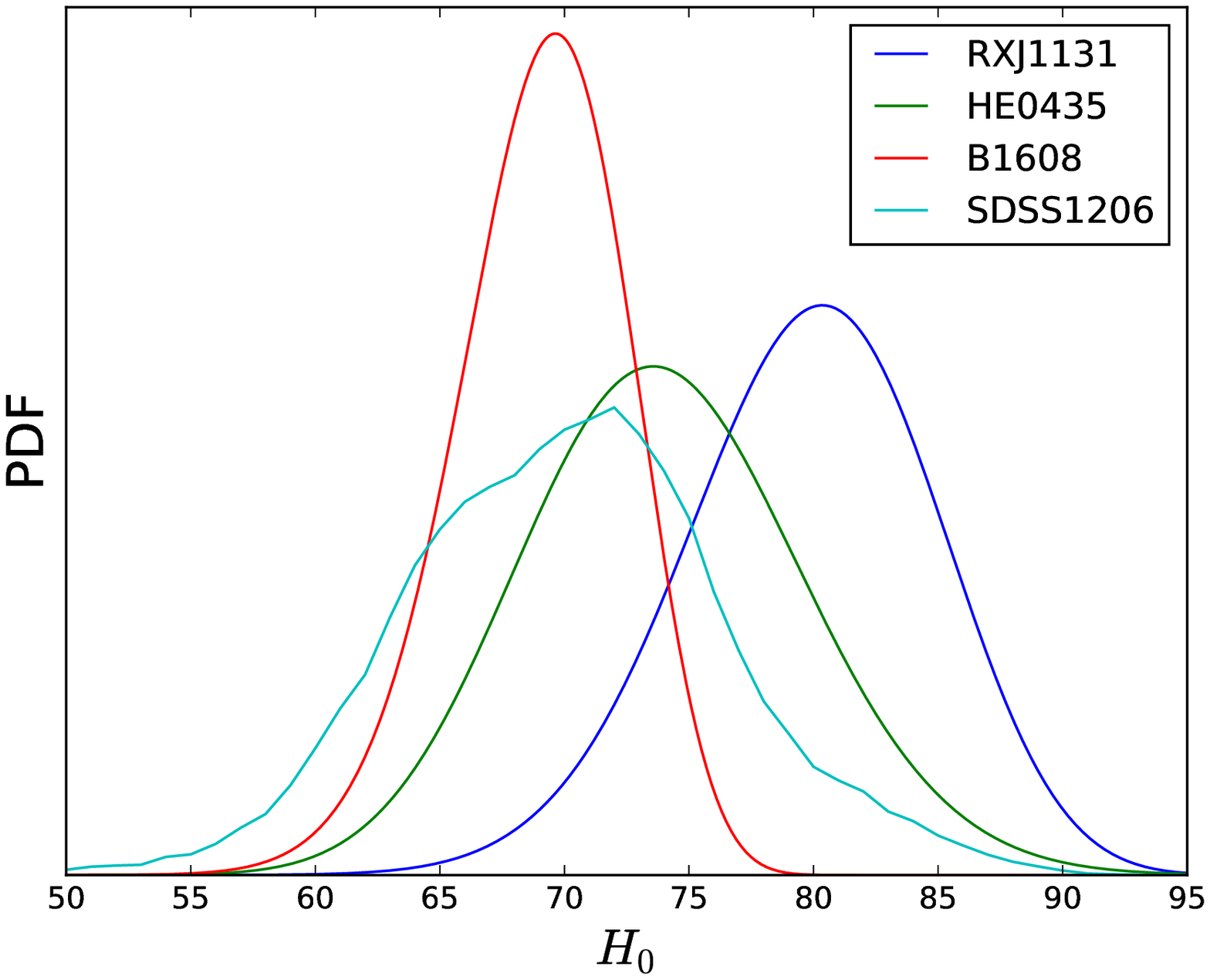}
    \caption{
    Left: Constraints on $H_0$ by combining 3 lenses out of the 4 lenses, excluding one. Right: Constraint on $H_0$ by each lens in the model independent cosmology.
    }\label{oneout}
\end{figure*}

To explore the potential impact of an outlier, we repeat our model independent
analysis using only
three of the time delay distances at a time, and investigate whether the
remaining system caused a shift in the final $H_0$ constraint.
Figure~\ref{oneout} shows the results. Removal of RXJ1131-1231 does
indeed have the greatest impact in alleviating tension with the Planck value
of $H_0$, and it was the lens system most in tension with the SNe reconstruction.
Note that as this is the lowest redshift system, one might expect that the SNe, which densely sample those distances, should provide an accurate result.
Finally, we can mention that independent measurement of the time delays
themselves shows a small, but interesting effect. The COSMOGRAIL team has been
extremely open and helpful about releasing the lightcurve monitoring data.
\citet{hojjati} used a GP method in 2013 to measure the time delay from
RXJ1131-1231 and found a value 1.4\% larger than COSMOGRAIL's (still within
$1\sigma$). H0LICOW
\citep{suyupriv} kindly put this value into their analysis pipeline and found
this decreased the value of $H_0$ from this system by 1.4\% (The magnitudes
don't need to be the same because changing $\Delta t$ also changes the lens
modeling.) So at least at a minor level, the values could
potentially come into greater agreement.


\section{Conclusion and discussions}

We carry out a model independent analysis (with no distance parametrizations or assumptions about any dark energy model)
using Gaussian process regression to determine the Hubble constant $H_0$ by anchoring the Pantheon Type Ia supernovae with time delay distances from the four publicly released, robust H0LiCOW lenses.
The analysis is done for both flat and non-flat universes.

Furthermore, we explore both the internal consistency of the
time delay systems estimations of $H_0$ and the external consistency
of the time delay distances with the GP expansion history derived
from SNe. This includes new probability distribution functions
for $H_0$ for each lens system in the model independent analysis.
No statistically significant tensions are found, although there
are some $\sim1\sigma$ trends that could be checked with future
data.

Our model independent results are $H_0=72.2\pm2.1\,$km/s/Mpc for
a flat Universe and $73.0^{+3.0}_{-2.8}$ km/s/Mpc allowing curvature. These
are consistent with the time delay and time delay plus SNe results
of \citet{Wong2019,Taubenberger2019,Collett2019}, made within
specific cosmological models or with polynomial fitting of the
distance relation. \citet{Collett2019} find $H_0 = 74.2^{+3.0}_{-2.9}$ km/s/Mpc for a flat Universe and $H_0 = 75.7^{+4.5}_{-4.4}$ km/s/Mpc allowing for curvature. Despite not assuming a specific model, the
uncertainties in our constraints are comparable to these, while
reducing possible bias. They are
also consistent with local distance measures of $H_0$, lying
midway between \citet{Freedman1907.05922} and \citet{2019ApJ...876...85R}. The strong lensing time delay plus SNe
model independent method looks quite promising as further data on
strong lens time delay systems becomes available.

For future studies, current surveys like the Dark Energy Survey (DES)~\citep{Treu2018} and the Hyper SuprimeCam Survey (HSC)
\citep{More2017}, and
the upcoming surveys like the Large Synoptic Survey Telescope (LSST) \citep{Oguri2010} and Euclid and WFIRST satellites
\citep{Barnacka1810.07265,Petrushevska} will bring us
thousands of lensed quasars and over one hundred
lensed SNe Ia, a part of which will have well-measured time delays~\citep{Liao2015}. With high-quality ancillary observations, some dozens
of systems will give us time delay distances at percent levels. Moreover, the angular diameter distances will be well-measured as well~\citep{Jee2016,Liao2019}.
In combination of these two kinds of distances, the Hubble constant could be constrained to sub-percent precision. However, at that stage,
systematic errors should be important. Dedicated analysis should be applied to individual lenses such that the combinations are robust.

Supernovae data will continue to improve as well, playing an
important role as a dense sampler of cosmic expansion history over
a wide range of redshifts. In addition to strong lensing, further local distance measurements such as gravitational waves from mergers of binary stars as standard sirens could
join this method and provide strong anchoring ability to model-independently determine the Hubble constant.

\section*{Acknowledgments}

We thank Stefan Taubenberger for providing the results for comparison. KL and AS would like to thank Zong-Hong Zhu and Beijing Normal University for the hospitality where the discussions on this project initiated. KL was supported by the National Natural Science Foundation of China (NSFC) No. 11603015
and the Fundamental Research Funds for the Central Universities (WUT:2018IB012). AS would like to acknowledge the support of the Korea Institute for Advanced Study (KIAS) grant funded by the Korea government.
EL is supported in part by the Energetic Cosmos Laboratory and by the U.S.\ Department of Energy, Office of Science, Office of High Energy Physics, under Award DE-SC-0007867 and contract no. DE-AC02-05CH11231.


\clearpage

\end{document}